\documentstyle[multicol,prb,aps,epsfig]{revtex} 
\tighten       

\begin{document}
\draft 
\title{Field-domain spintronics in magnetic semiconductor multiple quantum wells}
\author{David S\'anchez,$^{1,2,3}$ A.H. MacDonald,$^{2,3}$ and Gloria Platero$^1$}
\address{$^1$Instituto de Ciencia de Materiales de Madrid (CSIC),
Cantoblanco, 28049 Madrid, Spain \\
$^2$Department of Physics, Indiana University, Bloomington, Indiana 47405 \\
$^3$Department of Physics, The University of Texas at Austin, Austin, Texas 78712}
\date{\today}
\maketitle
\begin{abstract}
We develop a theory of non-linear growth direction transport 
in magnetically doped II-VI compound semiconductor multiple-quantum-well systems.  
We find that the formation of electric field domains can be controlled
by manipulating the space dependence of the band electron spin-polarization,
using its exchange coupling to local moments.  We emphasize the importance of 
band electron spin relaxation in limiting the strength of these effects.
\end{abstract}


\begin{multicols}{2}
\narrowtext

\setcounter{equation}{0}

\section{Introduction}
\label{sec-intro}
The giant magnetoresistance effect in magnetic metal multilayers\cite{bai88} occurs
because of the coupling of external magnetic fields to band electron spins through
their collective spin polarization.  The utility of this effect for information storage 
and field sensing 
devices has increased interest in exploring related spin-dependent transport properties in 
both ferromagnetic and paramagnetic semiconductors.  At the same time,
progress in the homo and hetero epitaxy of magnetically doped semiconductors
is creating new possibilities for engineered material geometries in which
new spin-dependent transport effects are likely to occur.~\cite{ans98,ohn98}
Large band electron spin polarizations can occur\cite{aws99} in diluted magnetic
semiconductors (DMSs),\cite{dmsreviews} for example in
II-VI compounds with Mn substituted on the group II site.
Among the striking phenomena already demonstrated in Mn doped II-VI quantum structures
are magnetically tunable quantum well barriers and interwell couplings,~\cite{kos87}
spin-dependent dynamics of polarized excitons in spin superlattices,~\cite{smy93}
optically-probed spin coherence,~\cite{kik97}
and the injection of highly polarized spin currents
into GaAs/AlGaAs light emitting diodes.~\cite{fie99}
In addition, $n$-doping of wide-gap II-VI magnetic semiconductor 
quantum wells (Zn$_{1-x-y}$Cd$_x$Mn$_y$Se)
has been achieved,\cite{smo97} yielding two-dimensional electron gas (2DEG)
systems that are exchange coupled to magnetic ions.~\cite{smo96}
The exchange interaction, $J_{sd}$, between $s$-electrons in the conduction band and 
Mn$^{++}$ $S=5/2$ local moments, results in band spin splittings larger than $\hbar \omega_c$,
the Landau level splitting. The spin-splittings can reach values as high as $20$~meV.\cite{cro00}
In fact, complete spin polarization can be achieved in quantum wells 
at relatively low magnetic fields ($\sim 1$~T).~\cite{cro95}

Although the study of electronic transport properties in DMS heterostructure systems is
still in its initial stages, interesting predictions have already been made.~\cite{egu98,guo01}
The present study is motivated in part by the recent growth of a modulation-doped ZnSe/(Zn,Cd,Mn)Se
multiple quantum well (MQW) system.\cite{ber00} In non-magnetic MQW systems, growth direction
transport phenomenology is enriched by an interplay between charge accumulation and 
resonant inter-well tunneling effects that results in the formation of 
electric field domains.  In this paper we report on a theory of the influence of  
exchange coupling with Mn ion spins on electric field domain formation and on the 
sensitivity of this influence to spin-relaxation rates within the quantum wells.

The formation of electric field domains is the hallmark
of dc-biased transport in weakly-coupled semiconductor superlattices.
Spontaneously generated inhomogeneities in the
spatial distribution of voltage drops, were proposed as an explanation for  
overall conductance ($G(V)$) oscillations,~\cite{esa70} discovered in 
pioneering growth direction MQW transport studies.
This early hypothesis was later confirmed by 
direct photoluminiscence measurements.~\cite{gra91}
Subsequently, highly doped GaAs/AlGaAs superlattices that
present sawtooth-like current--voltage ($I$--$V$) characteristics in the
negative differential conductance (NDC) region were studied in detail.
Along branches of the sawtooth two approximately constant electric field regions develop in
the sample, separated by a layer of accumulated electrons. 
In the following, we follow common usage in referring to these
layers with higher 2D electron density as \emph{monopoles}.
This non-equilibrium configuration enables resonant tunneling between ground and 
excited subbands in the high field region, minimizing the total resistance of the superlattice.
Increases in external voltage in this regime lead to sharp decreases in current,
followed by discrete jumps of the monopole region 
from a well to its upstream neighbor,
extending the high field domain over an additional period and
increasing the current. 

Past work has studied the dependence of domain formation and evolution on 
magnetic fields applied along the growth direction and on far-infrared 
radiation wavelength.
In the former case, the formation of Landau levels and scattering between them
introduces a new voltage scale for domain formation.~\cite{sch98}
In the latter case, photonic sidebands sustain the formation of the electric field
domains.~\cite{agu98}
The inclusion of the electronic spin in the study of perpendicular transport in
MQWs can also be expected to alter electric field domain formation physics, and 
exchange coupling to Mn spins should make it possible to tune these effects with 
relatively weak external magnetic fields.  This is the possibility that we explore 
at greater length in this paper.
In all these cases the electron-electron interaction, although small in the ground state 
in comparison with
typical energies of the system, can not be neglected since it is the Poisson equation
relating charge accumulation to field variations that is at the heart of field-domain formation.
For example it permits the experimentally observed\cite{kas94} multistability
of distinct stationary physical states at a fixed bias voltage.
Nonetheless, the mean-field Hartree approximation is sufficient to capture this 
physics in typical samples.
The non-linearity of the current versus voltage relationship between neighboring 
quantum wells, coupled with the non-locality of electron-electron interaction
effects leads to transport equations that can be solved only numerically, and
also to results that are sometimes difficult to interpret.

In this paper we deal with the formation of electric field domains in II-VI MQW systems
with one or more (II,Mn)VI quantum wells.
The main ingredients of our self-consistent theoretical model are:
(i) a theory for the tunneling current between two spin-polarized 2DEGs;
(ii) a continuity equation that 
accounts for relaxation of non-equilibrium spin populations;
(iii) a relationship between the up and down chemical potentials and their densities; 
(iv) the application of simple Hartree mean field theory to account for the Coulomb interaction;
and (v) a mean-field theory for the interaction between 2DEG electrons and Mn spins whose 
average polarization is very sensitive to external magnetic fields at low temperature.
We shall demonstrate that new features appear in the $I$--$V$ curve that depend on temperature
and Mn spin concentration, and explain why spin bottlenecks turn out to have a strong influence
in the instability regions.

The paper is organized as follows. In Section~\ref{sec-model} our theoretical model
is thoroughly explained. Sec.~\ref{sec-tech} is devoted to a discussion of 
technical details important for the numerical integration of the rate equations
that describe the time dependent charge, spin, and current distributions.
In Sec.~\ref{sec-res} we give numerical results and discuss their interpretation.
Finally, Sec.~\ref{sec-con} contains our conclusions.

\section{Theoretical Model}
\label{sec-model}
Following sucessful\cite{dmsreviews} early work on bulk systems 
by Kossut,\cite{kos76} Bastard,\cite{bas78} and Gaj,\cite{gaj78}
we account for the presence of Mn ions in DMS quantum wells,
by combining a phenomenological exchange model with a virtual crystal approximation
and mean-field-theory. 
The lattice parameters and the band Hamiltonian parameters of a II-VI heterostructure are 
assummed to change smoothly as Mn$^{++}$ spins are introduced in the system and
a $S=5/2$ quantum spin is assumed to be added to the low energy degrees of freedom
for each Mn spin.  The band electron system and the local moments are coupled 
by a ferromagnetic exchange interaction that favors 
parallel alignment of the local moment and band electron spins.
The total Hamiltonian of the system is:

\begin{equation}
{\cal H}={\cal H}_0+{\cal H}_{\rm T}+{\cal H}_{\rm scatt}+{\cal H}_{\rm int}^{ss}
+{\cal H}_{\rm int}^{sd}+{\cal H}_{\rm int}^{dd}+{\cal H}_{\rm sf}\,.
\label{eq-ham}
\end{equation}

The first four terms in the right-hand side of the equation
describe a conventional superlattice system with many weakly coupled
quantum wells:

\begin{enumerate}
\item ${\cal H}_0$ is the Hamiltonian for independent electrons in $N$ \emph{isolated}
quantum wells. Its energy spectrum is purely single-particle-like, and the 
quasiparticle spectrum is that of an isolated 
quantum well 2DEG ${\cal E}_j(\vec k_{||})=E_j+\xi(\vec k_{||})$,
where $\vec k_{||}$ is the wave vector parallel to the MQW heterointerfaces,
$\xi(\vec k_{||})=\hbar^2 k_{||}^2/2 m^*$, $m^*$ is the effective mass,
and $j$ is the quantum well subband index.
We will take $\vec k_{||}$ to be a continuous index , disregarding Landau-level
formation in the weak magnetic fields we will consider.
\item ${\cal H}_{\rm T}$ contains the tunneling amplitudes that couple quasiparticles in different
quantum wells. In weakly-coupled superlattices it is a good approximation to treat this term 
by leading order perturbation theory
as we discuss in Sec.~\ref{sec-tun}. 
\item ${\cal H}_{\rm scatt}$ contains the scattering terms within a quantum well that
allow a non-equilibrium quasiparticle to relax its excess energy, but does not contain
terms that permit the quasiparticle system to bring its spin-subsystems into
equilibrium. (These terms are absorbed in ${\cal H}_{\rm sf}$.) 
Because it is difficult to describe these scattering processes accurately,
or even to know what they are in particular systems, we will use a phenomenological
relaxation time approximation.  The time scale associated with these processes is typically 
rather short ($\tau_{\rm scatt}\sim 0.4$~ps).~\cite{smo96}
\item ${\cal H}_{\rm int}^{ss}$ is the electron-electron interaction in the conduction band
for which we will use a Hartree mean-field approximation. (See Sec.~\ref{sec-elec}).
\end{enumerate}

The remaining terms in Eq.~(\ref{eq-ham}) describe \emph{spin}-related physics:

\begin{enumerate}
\item ${\cal H}_{\rm int}^{sd}$ is the exchange interaction between $s$ conduction band 
electrons and Mn local moments, an interaction that turns out to be ferromagnetic in II-VI MQWs.
When the mean-field and virtual crystal approximations are employed, the effect of this coupling
is to make the subband energies spin-dependent in those quantum wells that contain Mn ions:
$E_j \rightarrow E_j^\sigma$.
\item ${\cal H}_{\rm int}^{dd}$ represents the antiferromagnetic super exchange 
interaction between Mn spins on neighboring lattice sites that has been found
to be important in modelling bulk DMS systems.\cite{aws99}
Since our intention here is to address the qualitative physics of field-domains in 
DMS MQW systems, we neglect ${\cal H}_{\rm int}^{dd}$.  We do expect, however, that these interactions 
will be important for detailed modelling of specific experimental systems.
\item ${\cal H}_{\rm sf}$ contains the microscopic processes that allow equilibrium to be 
established between spin subsystems within a quantum well.  
The fact that spin relaxation can be quite slow in the 
conduction band\cite{fla00} is one of the motivations for this work. Relaxation times 
in excess of $1$~ns have been established experimentally~\cite{kik97}
in II-VI semiconductor QWs without Mn. In II-VI DMS QWs these times are reduced
to tens of picoseconds (but still larger than $\tau_{\rm scatt}$).\cite{aws99}
We discuss the role of these terms at greater length in~\ref{sec-spin}.
\end{enumerate}

\subsection{2D-2D Tunneling}
\label{sec-tun}
The standard theory of tunneling relates the electric current between weakly-coupled subsystems
to  tunneling matrix elements and subsystem spectral functions.~\cite{mah00,zhe93}
In our case we will apply this theory to describe the current flowing between one
quantum well and its neighbor.
Since elastic and inelastic scattering times in the quantum wells are shorter than any other time
scale of the problem, we can follow the standard lines of tunneling theory and 
assume that the electrons in each well are in quasi equilibrium between
succesive tunneling events and that their temperature is that of the lattice.
We ignore interwell spin-flip processes, so that currents are carried between wells
by the two spin subsystems~\emph{in parallel}.
Accordingly, the current per spin from the $i$th well
to the ($i+1$)st well is given by the following general expression:

\begin{eqnarray}
J_{i,i+1}^\sigma=\frac{e \nu_0}{2\pi\hbar}
\sum_{\vec k_{i}\vec k_{i+1}} T_{\vec k_{i}\vec k_{i+1}}
\int d\varepsilon \, A_{\vec k_{i}}^\sigma(\varepsilon)
A_{\vec k_{i+1}}^\sigma(\varepsilon+eV_i)\nonumber \\
\times\left[f(\varepsilon-\mu_i^\sigma)-f(\varepsilon-\mu_{i+1}^\sigma+eV_i) \right] \,,
\label{eq-curr1}
\end{eqnarray}
where $\sigma=\left(\uparrow,\downarrow\right)$
is the conduction electron spin index,
$\nu_0=\frac{m^*}{2 \pi \hbar^2}$ is the 2D density of states per spin,
$T_{\vec k_{i}\vec k_{i+1}}$ is the transmission coefficient between particular 
wavevector states in the two quantum wells,
$eV_i$ is the voltage drop across the $i$th barrier and
$f(x)=1/[\exp (x/k_BT)+1]$ is the Fermi factor.
$\mu_i$ denotes the chemical potential in well $i$ measured from the bottom of well $i$.
A commonly used Lorentzian-shape function is chosen to represent the influence of
disorder on quasiparticles in 
the $j$th subband within the $i$th quantum well:

\begin{equation}
A_{\vec k_i}^\sigma(\varepsilon)=\frac{1}{\pi}\frac{\gamma}
{\left(\varepsilon-{\cal E}_j^\sigma(\vec k_i)\right)^2 +\gamma ^2} \,.
\end{equation}
This form for the spectral function results from neglecting the real 
part of the disorder self energy, which introduces an unimportant rigid shift of the quasiparticle
energies, and the energy dependence of its imaginary part.
$\gamma$ is treated as a phenomenological parameter whose value may vary substantially
from sample to sample and is to be taken from experiment.
($\gamma = \hbar/2 \tau_{\rm scatt}\sim 1$~meV).
In weakly-coupled superlattices the broadening due to scattering is much larger than
the miniband width so that tunneling between quantum wells is 
\emph{sequential} rather than band like. 
In this regime an electron undergoes many scattering events in one well 
before tunneling to the next well.
Because of the epitaxial nature of the samples in question, we assume that the tunneling process
conserves parallel momentum, {\it i.e.}, effects such as interface roughness are not taken into account.
This approximation is made for the sake of definiteness and does not influence the 
qualitative conclusions we will reach.
In addition, the typical electronic densities in particular quantum wells
are assumed to be $\sim10^{11}$cm$^{-2}$ so that only
the lowest subband is appreciably populated in the quasi-equilibrium state of a quantum well.
Therefore, when current flows dominantly by a transition from the ground state of well $i$ to the
first excited state in well $i+1$, rapid relaxation to the lowest subband 
(via, e.g., emission of a LO phonon) is assumed.
This simplifications lead us to an analytical expresion
for Eq.~(\ref{eq-curr1}) at $T=0$~K:

\begin{eqnarray}
J_{i,i+1}^\sigma=\frac{e\nu_0}{2\pi ^2\hbar}
\,\Xi(\mu_i^\sigma,E_{i\,1}^\sigma,\mu_{i+1}^\sigma,E_{i+1\,1}^\sigma,eV_i) \nonumber \\
\times \sum_j T_{j} \frac{2\gamma}{(E_{i\,1}^\sigma-E_{i+1\,j}^\sigma+eV_i)^2+(2\gamma)^2} \,.
\label{eq-curr2}
\end{eqnarray}

The sum in Eq.~(\ref{eq-curr2}) is extended over all subands in the $(i+1)$st well.
The transmission coefficients from the ground subband to the $j$th subband of the neighbouring well,
$T_{j}$, are calculated by means of the transfer Hamiltonian method.\cite{agu97}
Since this approach involves only orbital degrees of freedom,
$T_{j}$ does not depend explicitly on $\sigma$.
We are therefore implicitly restricting ourselves to the case when
the spin splittings are not so large as to mix different subbands.
The function $\Xi$ expresses the width of the energy ``window'' available for tunneling.
For tunneling between the lowest subbands:

\begin{displaymath}
\Xi \equiv \left\{ \begin{array}
     {r@{\quad:\quad}l}
\mu_{i}^\sigma-E_{i\,1}^\sigma & \mu_{i+1}^\sigma-eV_i<E_{i\,1}^\sigma \\
-\left(\mu_{i+1}^\sigma-E_{i+1\,1}^\sigma\right) & \mu_{i}^\sigma+eV_i<E_{i+1\,1}^\sigma \\
\mu_i^\sigma-\mu_{i+1}^\sigma+eV_i & \mbox{otherwise} \,, \end{array} \right.
\end{displaymath}
where $E_{i\,1}$ is the first subband energy in the $i$th-well and so on.
The expressions for the energy window for tunneling to higher subbands are similar
and differ only through the absence of Pauli blocking effects in the target layer.

\subsection{Spin splitting and spin relaxation}
\label{sec-spin}
The exchange interaction between the conduction band electrons and the Mn$^{++}$ ions produces
a giant spin splitting of the 2DEG even in the presence of a small magnetic field.
The bottom of the band for each spin rigidly shifts accordingly:

\begin{eqnarray}
E_{j}^\sigma &=& E_{j} - s \Delta \\
\Delta &\equiv&  J_{sd}N_{\rm Mn}SB_S \left( \frac{g\mu_BBS}{k_B T_{\rm eff}}\right)\,,
\label{eq-delta}
\end{eqnarray}
where $E_{j}$ is the $j$th miniband center,
$s=+(-)$ for $\sigma=\uparrow$($\downarrow$),
$J_{sd}$ is the exchange integral, $N_{\rm Mn}$ is the 
density of Mn$^{2+}$ ions with spin equals to $S=5/2$,
$B_S$ is the Brillouin function, and $T_{\rm eff}$ is an effective temperature
which can, in general, include a correction due to antiferromagnetic interactions
between neighboring Mn ions.\cite{aws99}
We have assumed here that the magnetic field ($B$) orientation is
such that the quasiparticle energy is lowered for \emph{up} spins.
Note that our virtual crystal approximation for the 
Mn ions implies that the mean-field experienced by 
band electrons of spin $\sigma$ is spatially uniform.  
In this paper we take the field experienced by the local moments in Eq.~(\ref{eq-delta})
to be the external magnetic field.  In fact, the mean-field approximation we employ
can be extended\cite{lee00} to include the contribution of spin-polarized band electrons to
the total effective magnetic field experienced by the local moments.  When this is done,
ferromagnetism results. That extention of the models and approximations employed
here does in fact appear to account for the main features of the carrier induced ferromagnetism
that occurs\cite{die00,koe00,ourreview} in doped diluted ferromagnetic semiconductors.
The transition temperatures in these systems can be
substantial\cite{ohn99} for p-type systems and for higher carrier densities.  In the case of 
relatively low-density n-type systems, the ferromagnetic transition temperatures 
will be low and ferromagnetism does not necessarily occur,
when direct interactions between the Mn local
moments are included in the theory.  Experimentally, most n-type DMS 2DEGs show no
evidence for ferromagnetism.  In this paper we will assume that the contribution of
the spin-polarized conduction band system to the effective field experienced by the Mn
local moments is negligible.  If ferromagnetism did occur, the non-linear field-domain
transport physics we discuss in this paper would be further enriched.

In modelling spin-relaxation within a quantum well we 
start by nelecting the transport currents $J_{i,i+1}^\sigma$.~\cite{mac99}
For instantaneous spin-splittings smaller than the Fermi
energy we use the following phenomenological spin-relaxation rate equation
within each quantum well:

\begin{equation}
\frac{dn_i^\sigma}{dt}=-\frac{\mu_i^\sigma-\mu_i^{\overline{\sigma}}}
{\tau_{\rm sf}}\nu_0 \,,
\label{eq-rate1}
\end{equation}
where $\overline{\sigma}$ is the spin opposite to $\sigma$ and
$\tau_{\rm sf}$ is the spin-scattering time.
(We neglect the dependence of $\tau_{\rm sf}$ on $\Delta$, which
may be important in certain samples\cite{cro95}).
This equation implicitly assumes the linear non-interacting 2DEG 
relationship between density and chemical potential, so that 
correlation effects are not taken into account:\cite{mac00}
\begin{equation}
n_i^{\sigma}=\nu_0(\mu_i^{\sigma}- E_{i\,1}^{\sigma} ).
\label{eq-ni}
\end{equation}
In the absence of a driving bias voltage, the solution of Eq.~(\ref{eq-rate1})
at $t\to\infty$ is
$n_i^\uparrow = \left( n_i + \nu_0\Delta \right)/2$ 
and $n_i^\downarrow = \left( n_i - \nu_0\Delta \right)/2$.
Here $n_i=n_i^\uparrow+n_i^\downarrow$ is the total density of carriers in the $i$th well.
We can see from this asymptote that Eq.~(\ref{eq-rate1}) is valid only for spin splittings smaller
than the chemical potential.

\begin{figure}[ht]
\centerline{
\epsfig{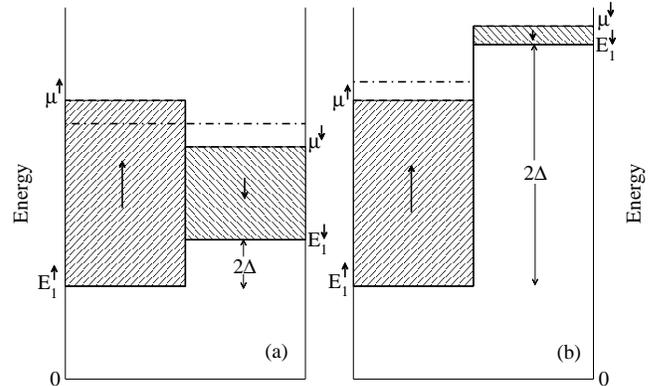}
}
\caption{Schematic illustration of spin-relaxation within a quantum well.
The dashed lines indicate initial non-equilibrium chemical potentials 
while the dot-dashed line the equilibrium ($t\to\infty$)
chemical potentials.  Panel (a) is for the case of spin-splittings
smaller than $\mu^\uparrow-E_1^\uparrow$ while panel (b) is the case of 
for spin-splittings larger than $\mu^\uparrow-E_1^\uparrow$ for which
the equilibrium state is completely spin polarized.
The zero of energy in these plots is the electrostatic potential of
the quantum well.}
\label{fig1}
\end{figure}

For $\Delta$ greater than the chemical potential, Eq.~(\ref{eq-rate1}) must be modified
using Eq.~(\ref{eq-ni}):

\begin{mathletters}
\label{eq-rate2}
\begin{eqnarray}
\frac{dn_i^\downarrow}{dt} & = & -\frac{n_i^\downarrow}{\tau_{\rm sf}}
= -\frac{\mu_i^\downarrow-E_{i\,1}^\downarrow}{\tau_{\rm sf}}\nu_0  \\
\frac{dn_i^\uparrow}{dt} & = & - \frac{dn_i^\downarrow}{dt} \,.
\end{eqnarray}
\end{mathletters}
For large enough $\Delta$, Eq.~(\ref{eq-rate2}) leads to an equilibrium state
with full spin polarization. (See Fig.~\ref{fig1}).

Adding transport currents to these considerations leads to the following
discrete continuity equations for the spin population in each quantum
well:
\begin{equation}
\frac{dn_i^\sigma}{dt}=\frac{J_{i-1,i}^\sigma-J_{i,i+1}^\sigma}{e}
-\frac{\mu_i^\sigma-\mu_i^{\overline{\sigma}}}{\tau_{\rm sf}}\nu_0 \qquad i=1,\ldots,N
\label{eq-rate3}
\end{equation}
for the case $\mu_i^\uparrow-E_{i\,1}^\uparrow>2\Delta$.
Otherwise, Eq.~(\ref{eq-rate2}) must replace the second term on
the right-hand side of Eq.~(\ref{eq-rate3}).

\subsection{Electrostatics}
\label{sec-elec}
In large area heterostructures, the Coulomb interaction is usually included in a Hartree
mean-field approximation.
The Poisson equation relates the electrostatic potential drop across
MQW barriers, $V_i$, to the charge distribution among the quantum wells:

\begin{equation}
\frac{V_i-V_{i-1}}{d}=\frac{e}{\epsilon}\left(n_i-N_w\right) \qquad i=1,\ldots,N\,.
\label{eq-poi}
\end{equation}
where $d$ is the superlattice period, $\epsilon$ is the sample average permittivity,
and $N_w$ denotes the doping density within the wells.
(Experimentally, doping is usually accomplished by 
placing a ZnCl$_2$ layer in the barrier layers; this difference in electrostatics
compared to our model has no important consequences.)~\cite{smo96}

By inspection of Eqs.~(\ref{eq-rate3}) and (\ref{eq-poi}) it is obvious that a
set of boundary conditions must be provided for $n_0$ and $n_{N+1}$.
Within our model these layers play the role of source and drain, respectively.
A simple way to represent source and drain~\cite{wac97} is to fix the density in both layers at high
values.  We take 
\begin{equation}
n_0,\,n_{N+1} \equiv \kappa N_w
\label{eq-bc}
\end{equation}
where $\kappa >1$ is an adjustable parameter.
More sophisticated models~\cite{agu97,bon00} have proven that a proper description of the contacts
can have a strong effect on the selection of the the transport
equation solution when multistability occurs,
especially when dynamical solutions are allowed.~\cite{bon00}
However, we choose not to delve into these effects in detail here since our main interest 
is on spin effects.
For the sake of definiteness, the contacts are taken to be unpolarized throughout our calculations;
including spin-polarized injection in our theory would be
straightforward and indeed this may be a very interesting avenue to explore in future experimental
and theoretical studies.
With this representation of the source and drain, fixing 
the overall bias voltage,
\begin{equation}
V=\sum_{i=0}^{N}V_i,
\label{eq-bias}
\end{equation}
closes the set of equations.

\section{Numerical Considerations}
\label{sec-tech}
Our model contains $5N+1$ unknown functions of time
($2N$ chemical potentials, $\mu_i^\sigma$, $2N$ electronic densities, $n_i^\sigma$,
and $N+1$ voltage drops, $V_i$).
These unknown functions are determined by the $2N$ constitutive equalities (Eq.~(\ref{eq-ni})),
$2N$ rate equations (Eq.~(\ref{eq-rate3})), $N$ Poisson relations (Eq.~(\ref{eq-poi})), and
the total bias condition (Eq.~(\ref{eq-bias})).
Thus, giving a physically sensible initial profile, the system of algebraic-differential equations
can be integrated to yield a definite solution.
Standard numerical methods are employed in solving Eq.~(\ref{eq-rate3}),
being careful to use the appropriate spin-relaxation equation
(either Eq.~(\ref{eq-rate1}) or Eq.~(\ref{eq-rate2}))
at each time step.

The total current density $J(t)$ traversing the sample at time $t$ is determined by the 
following procedure.
Differentiate Eq.~(\ref{eq-poi}) with respect to time and substitute the result into
the sum of Eq.~(\ref{eq-rate3}) over spin indices
in order to eliminate densities and chemical potentials.
This leads to the following current which has the same value when 
evaluated for any well index, $i$:

\begin{equation}
J(t)=\frac{\epsilon}{d}\frac{dV_i}{dt}+J_{i,i+1}(t)
\label{eq-amp}
\end{equation}

The first term of the right-hand member of the previous equation is the displacement current
whereas the second term is the tunneling current,
$J_{i,i+1}(t)=J_{i,i+1}^\uparrow(t)+J_{i,i+1}^\downarrow(t)$.
For static steady state solutions discussed in this paper only the latter term is finite. 

\section{Results}
\label{sec-res}
We focus on ZnSe/Zn$_{1-x-y}$Cd$_x$Mn$_y$Se DMS MQWs.
A value of $x\sim 0.2$ has been chosen to be consistent with barriers
($\sim 200$~meV) sufficiently high to capture more than one subband in the quantum wells.
For definiteness, we focus on the case where Mn has been incorporated
only in the \emph{central} well of the superlattice.
In experimental samples the value of $y$ can be varied over a wide range.
We expect that the field-tuned field-domain effects we discuss will be 
strongest at 
moderate Mn densities, large enough to give rise to sizable spin-splittings 
but not so large as to increase the spin scattering rate excessively.

The experimental samples reported on in Ref.~\onlinecite{ber00} possess 
ZnSe barriers too thick ($\sim 30$~nm) for perpedicular transport. 
Under those conditions, the coefficients $T_j$ are so small that
electron tunneling would likely occur via
impurity channels in the barriers or
through different symmetry points in the host semiconductor band structure,
violating the assumptions of our theory.
For these model calculations we choose a smaller barrier width: $b=5$~nm.
The remaining phenomenological parameters have been fixed on the basis
of available experimental data:
$w=10$~nm, $m^*=0.16 m_e$, $N_w=2\times 10^{11}$~cm$^{-2}$, $\tau_{\rm sf}=10$~ps, and $\kappa=1.5$.
$w$ is the well width and $\kappa$ specifies the source and drain densities
(see Eq.~(\ref{eq-bc})).

In Sec.~\ref{sec-lin} we study self-consistent steady-state solutions for low voltages.
This will help us understand the more complicated behaviors
that result from instabilities in the NDC voltage regime.
(See Sec.~\ref{sec-ndc} below.)

\subsection{Linear regime}
\label{sec-lin}
\begin{figure}[ht]
\centerline{
\epsfig{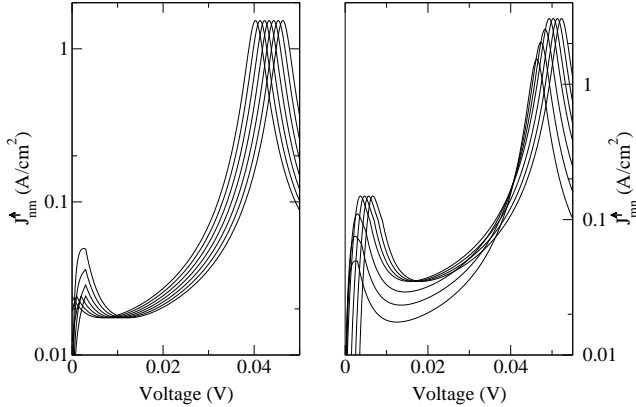}
}
\caption{(a) Current density flowing from a non-magnetic well to a magnetic one
for $\Delta$=0-6~meV in steps of 1~meV.
The rightmost curve corresponds to $\Delta$=0~meV.
(b) Same as (a) but the carriers are now flowing from the magnetic well to the
non-magnetic one. The leftmost curve is for $\Delta$=0~meV.}
\label{fig2}
\end{figure}

\begin{figure}[ht]
\centerline{
\epsfig{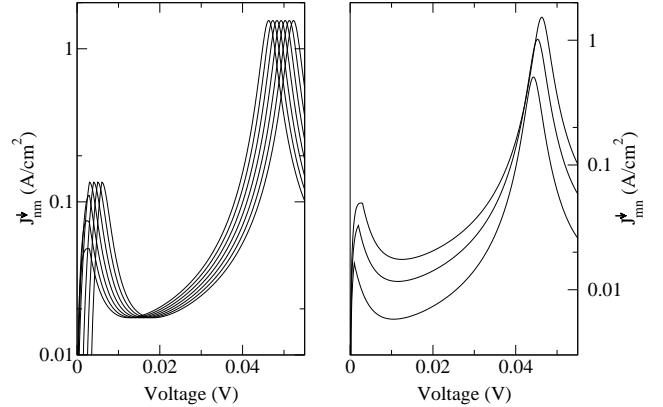}
}
\caption{(a) Same as Fig.~\ref{fig2}(a) for down spins.
The leftmost curve corresponds to $\Delta$=0~meV.
(b) Same as Fig.~\ref{fig2}(b) for down spins.
The rightmost curve is for $\Delta$=0~meV.}
\label{fig3}
\end{figure}

The behavior of the spin-dependent current density in Eq.~(\ref{eq-curr2}) depends on
many variables. In Figs.~\ref{fig2} and~\ref{fig3} we have plotted $J_{i,i+1}^\sigma$,
for the case where $\mu_m^\uparrow$
and $\mu_m^\downarrow$ have been set to their equilibrium values
($\mu_m^\uparrow$=$\mu_m^\downarrow$=$\mu_m$).
Thus the spin-dependent chemical potential, measured from the bottom of the well,
will depend on the value of $\Delta$
(see Fig.~\ref{fig1}).

Fig.~\ref{fig2}(a) plots the up-spin current from a \emph{n}on-magnetic well
to a \emph{m}agnetic one, $J_{nm}^\uparrow$, {\it i.e.}, net electron flow from a magnetic well
to an non-magnetic one. For $\Delta=0$~meV and low voltages the behavior is ohmic,
as expected. A larger bias results in the alignment of the ground states
of both wells within $\gamma$, giving rise to a first maximum in the current.
After this bias is exceeded, Eq.~(\ref{eq-curr2}) implies the appearance of a NDC region due
to subband mismatch and of a second peak at higher biases when the first subband is aligned with the
second subband of the next well.
Increasing $\Delta$ decreases the value of the first peak since the up-spin density in the magnetic
well increases and fewer states are available for tunneling.
The peak corresponding to tunneling from first$\rightarrow$second subband
is not affected in magnitude but its position is shifted to lower bias voltages
because the bottom of the up-spin subband goes down.
For a given value of the spin splitting ($\Delta\sim 3$~meV), the magnetic well is fully
polarized so that the first peak magnitude can no longer vary its value,
a displacement due to subband lowering is observed instead. 
In Fig.~\ref{fig2}(b) the up-spin current from a magnetic well
to a non-magnetic one is shown.
Here the first peak becomes larger as $\mu_m^\uparrow$ grows since more electrons take part
in the tunneling. This increase ceases once the fully polarized situation is achieved.
In addition larger voltages are needed to align the energy levels as $\Delta$ increases
and the peak moves to higher voltages. Field-domain physics is at heart
controlled by the interplay of electrostatics and the layer-to-layer non-linear $I$--$V$ relationships.
These examples illustrate that the layer-to-layer non-linear $I$--$V$'s can be altered by 
$\Delta$ (and hence an external magnetic field) when one of the layers contains Mn local
moments. It follows that the field-domain structure must be $\Delta$ and field dependent as 
we show below.

The corresponding down-spin current is plotted in Fig.~\ref{fig3}.
The considerations explained above for the up spin case may be invoked to understand this figure.
For $\Delta\gtrsim 3$~meV (see Fig.~\ref{fig3}(b))
the down-spin current flowing from the magnetic to the non-magnetic well
($J_{mn}^\downarrow$) vanishes for any value of the applied
bias since in the fully polarized regime,
no down-spin carriers are present in the magnetic well.

\begin{figure}[ht]
\centerline{
\epsfig{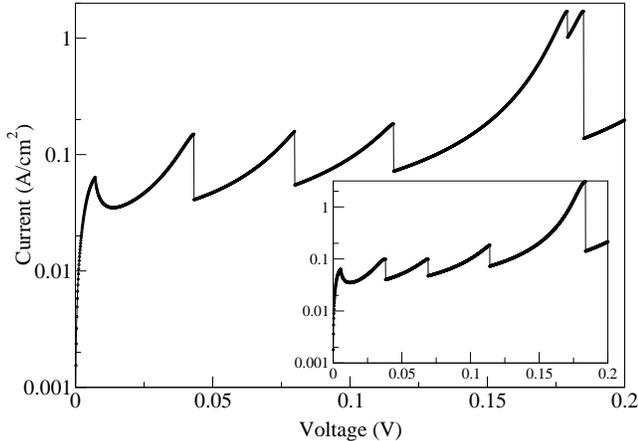}
}
\caption{$I$--$V$ characteristics for $\Delta=3$~meV.
The first peak occurs near where the first$\rightarrow$first subband resonant condition
occurs in a configuration with no field domains.
The following three branches reflect steady states with monopoles in one of the 
three quantum wells of the system.  The monopoles move upstream with increasing
bias voltage, increasing the number of high field barriers.
Note that the final peak in the I-V curve is split in the spin-dependent 
transport case.
Results for the $\Delta=0$~meV case are shown in the inset for comparison.
The vertical lines that connect different branches of the I-V curve are guides to the eye.
}
\label{fig4}
\end{figure}

A numerical calculation for $N$=3 gives rise to the MQW current--voltage characteristics
shown in Fig.~\ref{fig4}.
$\Delta$ has been set to 3~meV. At low voltages the behavior is linear, up to the peak which
marks alignment to within $\sim \gamma$ of the first subbands of all wells.
Then, after entering the NDC region, an increase of the current occurs because an electric field
domain has formed.
The underlying mechanism can be understood in the following terms.
As the voltage is increased, the system prefers 
to mantain regions of low and high electric fields where intra-subband and inter-subband 
resonant tunneling can occur, rather than maintaining a constant field.
The high field regime forms in the downstream side of the sample.
In this way, the superlattice minimizes the total resistance and current increases.
In order to build the inhomogenous field, according to Poisson equation~(\ref{eq-poi})
an excess electron density, or a monopole, must accumulate in the well at the boundary between low and 
high field regions.
As the voltage is further increased, configurations with particular monopole 
locations sucessively become unstable and the monopole moves by one quantum well to
increase the width of the high field region.  Monopole motion 
gives rise to an abrupt decrease of the current, which is followed by a gradual
increase as the voltage increases further.
These successive monopole position jumps lead ultimately to 
the sawtooth-like $I$--$V$ curve shape of Fig.~\ref{fig4}.
The inset in this Figure shows the corresponding $\Delta=0$~meV $I$--$V$ curve for comparison.
The most obvious change is the appearance of a new branch that develops close to
the first$\rightarrow$second subband transition (see below).

\begin{figure}[ht]
\centerline{
\epsfig{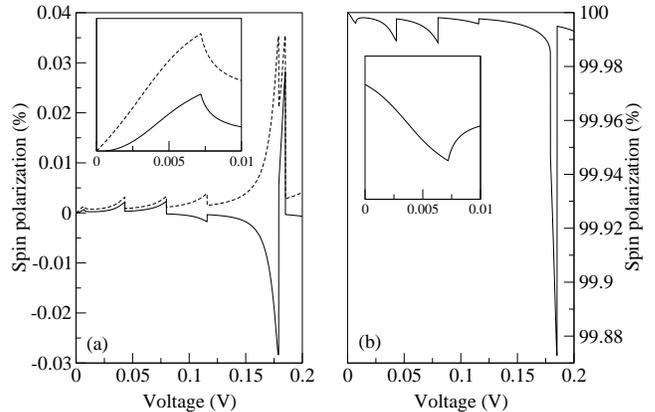}
}
\caption{(a) Spin polarization for the upstream (downstream) quantum well is represented by 
full (dashed) lines.
The inset highlights the low voltage regime. (b) Same as (a) but for the magnetic quantum well.}
\label{fig5}
\end{figure}

The somewhat complicated behavior of the bias voltage dependence of the 
steady state spin polarization
is depicted in Fig.~\ref{fig5} for $N=3$ and $\Delta=3$~meV.
Notice that in Fig.~\ref{fig5}(a)
a non-zero steady state spin polarization is induced in the non-magnetic wells
by the spin-dependent transport currents.
We have assumed here that the spin-relaxation time constant $\tau_{\rm sf}$ is the same in all
quantum wells (10~ps). In practice these times are likely to be considerably larger in the 
non-magnetic wells, and the induced spin-polarization in these wells will have an
even larger importance.
For low voltages, the inset of Fig.~\ref{fig5}(b) shows a reduction of the magnetic well
spin polarization. This effect can be understood by realizing that the minority spin current
from a non-magnetic well to a magnetic one, $J_{nm}^\downarrow$, grows
when the system is driven out of equilibrium by a small bias voltage.
This occurs also in $J_{nm}^\uparrow$ but at a smaller rate because of 
the spin subband displacement in the magnetic well
(compare Figs.~\ref{fig2}(a) and \ref{fig3}(a)).
Therefore, at low voltages the magnetic well polarization must decrease.
Correspondingly, the upstream-well polarization must increase (see the inset of Fig.~\ref{fig5}(a))
since down spins leak out of it.
The enlargement of the downstream-well spin polarization has the same origin.
(see Figs.~\ref{fig2}(b) and \ref{fig3}(b)).

\subsection{NDC regime}
\label{sec-ndc}
\begin{figure}[ht]
\centerline{
\epsfig{file=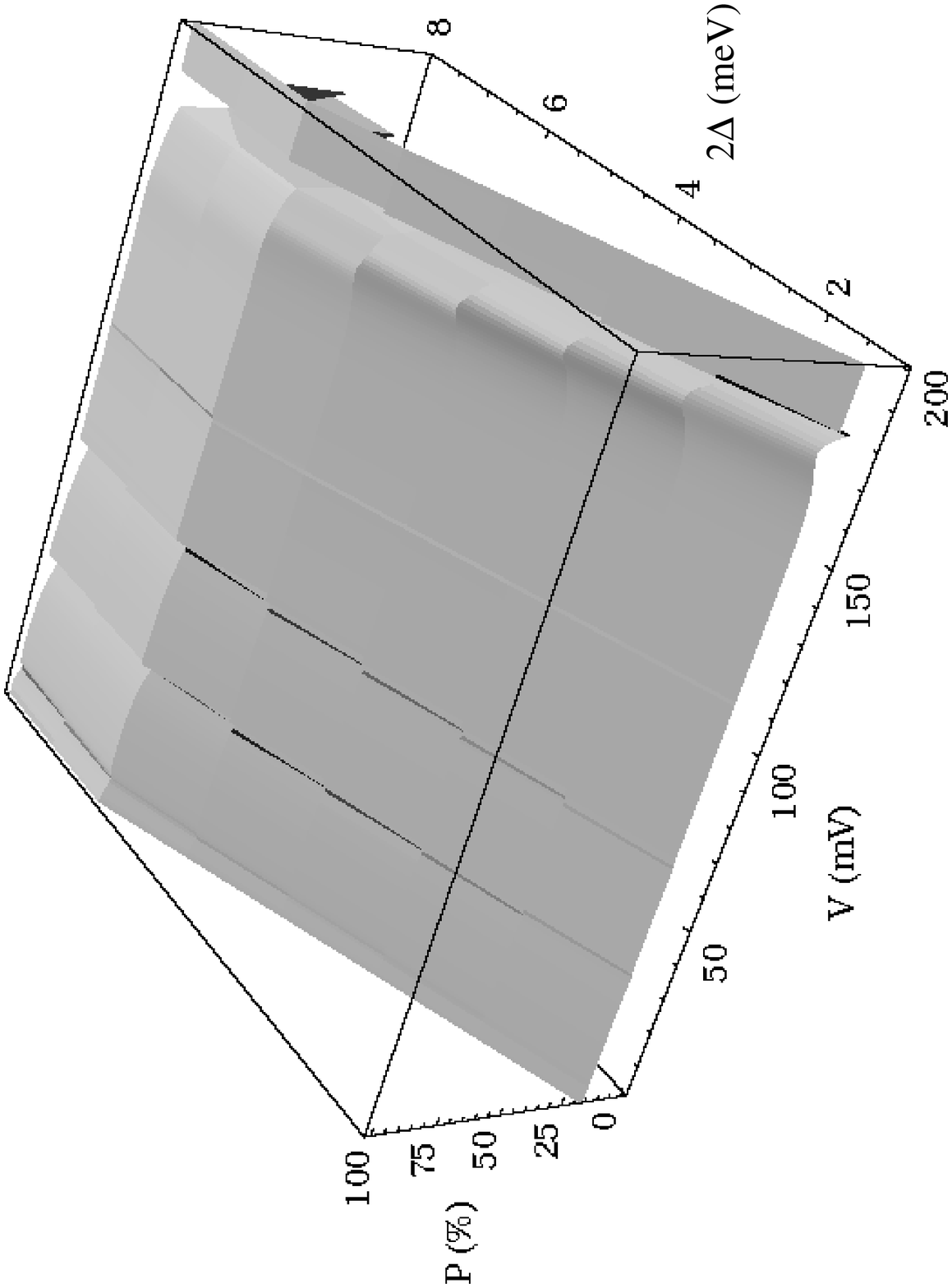,angle=270,width=0.47\textwidth,clip}
}
\caption{3D plot describing the dependence of the magnetic well spin polarization
on voltage bias and spin-splitting $\Delta$. In constructing this figure, 
$\tau_{\rm sf}$ has been increased to 10~ns
(still a reasonable value)~\protect\cite{kik97}
in order to magnify the effects commented on the text.}
\label{fig6}
\end{figure}

As explained above, once the bias voltage reaches the NDC regime, electric field
domains form in the sample.  As we now discuss, their formation strongly influences
spin-polarizations in both magnetic and non-magnetic quantum wells, with discontinuities
associated with every break in the $I$--$V$ curve.
The magnetic well polarization, $P$, varies particulary strongly, especially
when the monopole moves through the magnetic well, and becomes stronger as the 
spin-splitting is increased.  
This can be seen in Fig.~\ref{fig6}, which describes the general behavior of $P$
with voltage and spin splitting.

\begin{figure}[ht]
\centerline{
\epsfig{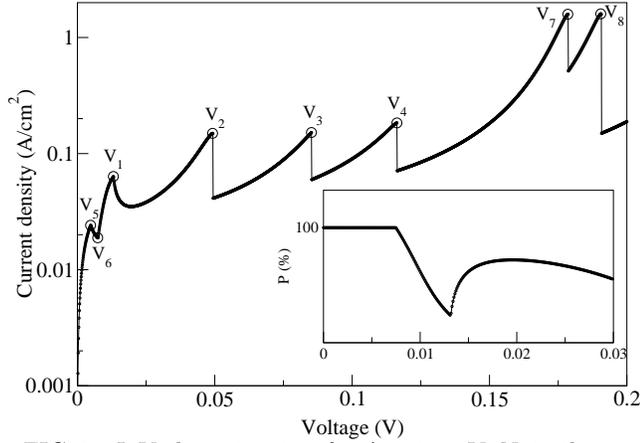}
}
\caption{$I$--$V$ characteristics for $\Delta=6$~meV. Note the appparance of new 
branches in the steady-state $I$--$V$ curve.
The spin polarization in the magnetic quantum well at low voltages is plotted in the inset.}
\label{fig7}
\end{figure}

If $\Delta$ is further increased ($\Delta$=6 meV),
new branches appear in the $I$--$V$ curve (see Fig.~\ref{fig7}).
The extra branch that appeared close to the $E_1\rightarrow E_2$ transition at smaller 
$\Delta$ is now fully developed.
In addition, a new branch forms \emph{before} the $E_1\rightarrow E_1$ transition.

\begin{figure}[ht]
\centerline{
\epsfig{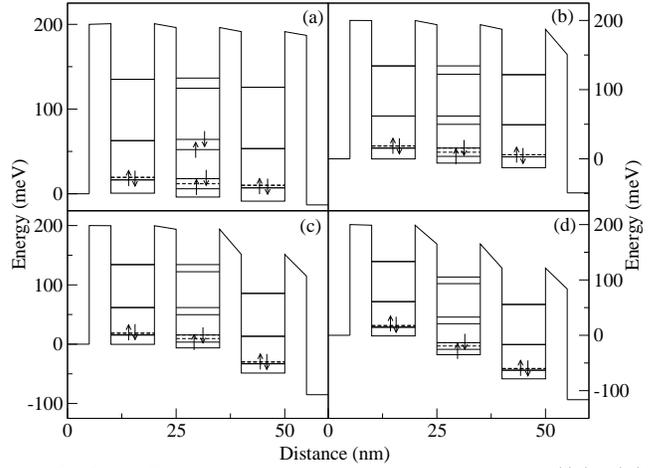}
}
\caption{MQW profiles for $V_1$, $V_2$, $V_3$, and $V_4$
((a), (b), (c), and (d) respectively) marked in Fig.~\ref{fig7}.
Resonant levels (chemical potentials) are depicted by solid (dashed) lines.
The spin polarizations plotted in Fig.~\ref{fig5}(a) are not observable
on the scale of these figures.}
\label{fig8}
\end{figure}

To better understand the subtle interplays that control these features,
we have studied the self-consistent steady-state well-dependent spin-polarizations at the
particular voltages marked in  Fig.~\ref{fig7}. 
The series of MQW electrostatic profiles illustrated in 
Fig.~\ref{fig8} are dominated by classical field-domain physics not qualitatively 
influenced by the spin-dependent nature of the transport.
Fig.~\ref{fig8}(a) ($V_1$) is the highest voltage at which intrasubband
($E_1\rightarrow E_1$) resonant tunneling can be maintained.
The electric field drops almost linearly along the system.
In Fig.~\ref{fig8}(b) ($V_2$) the formation of a high electric field domain in the last barrier
is clearly observed, favouring a resonant condition
between the third well and the collector first-excited subband.~\cite{note}
The second branch in the NDC region ($V_3$) involves
the generation of a larger high field domain (see Fig.~\ref{fig8}(c)). 
The domain wall is now located in the magnetic well.
A jump to the first well (Fig.~\ref{fig8}(d))($V_4$) is accompanied by further 
expansion of the high field domain.
In this situation all tunneling within the wells
takes place between the ground and the excited states,
followed by a rapid relaxation to the first subband.

\begin{figure}[ht]
\centerline{
\epsfig{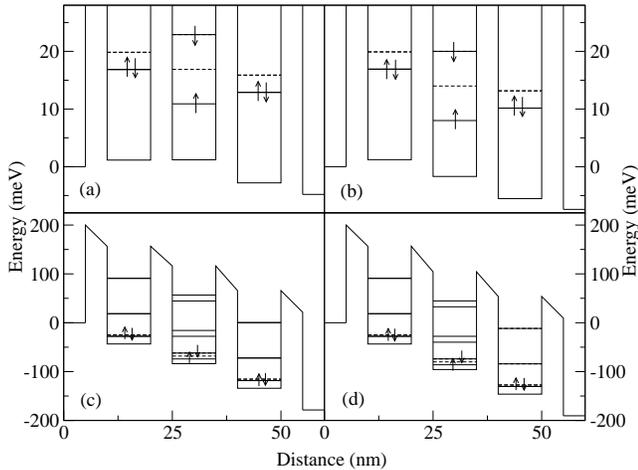}
}
\caption{Same as Fig.~\ref{fig8} for $V_5$, $V_6$, $V_7$, and $V_8$
((a), (b), (c), and (d) respectively).}
\label{fig9}
\end{figure}

Fig.~\ref{fig9} illustrates spin-populations near $I$--$V$ features where the 
spin-dependent element introduced by the magnetic quantum well plays a 
qualitative role.  
Fig.~\ref{fig9}(a) describes the position of energy levels
for the voltage $V_5$ marked in Fig.~\ref{fig7}.
The subband energies in the non-magnetic wells ($E^\uparrow$ and $E^\downarrow$) are quasidegenerate.
Notice that resonant tunneling occurs between $E_{1\,1}^{\uparrow}$
and $E_{2\,1}^\uparrow$. Further increase of the voltage, however, results in a decrease of
the current since now $E_{1\,1}^{\uparrow,\downarrow}$
is then off-resonance (see Fig.~\ref{fig9}(b)).
The current is then increased again since
$E_{1\,1}^{\downarrow}$ starts to match $E_{2\,1}^\downarrow$.
This explains why $P$ does not show the behavior
observed for smaller $\Delta$ ($\Delta$=3 meV) in the linear regime.
(In contrast to the inset of Fig.~\ref{fig5}(b), the inset of Fig.~\ref{fig7}
presents a flat polarization at low voltages.)
$P$ starts to increase only when $V$
is such that $E_{1\,1}^{\uparrow}$
reaches within $\gamma$ of $E_{2\,1}^\downarrow$.
The splitting of the branch at higher bias voltage where 
all transport occurs via intersubband 
($E_1\rightarrow E_2$) resonant tunneling has a similar explanation.
The Fig.~\ref{fig9}(c) is stabilized by alignment of $E_{1\,1}^{\uparrow}$
with $E_{2\,2}^\uparrow$ levels.
Subband mismatch at higher bias gives rise to a sharp reduction of the current,
which increases later, as the voltage is increased and resonant tunneling between
$E_{1\,1}^{\downarrow}$ and $E_{2\,2}^\downarrow$ levels is achieved.
Incidentally, the dips in $P$ illustrated in Fig.~\ref{fig5} can be explained in this way.
When the latter alignment occurs, a large flow of down-spin carriers streams
towards the magnetic well, causing a sharp decrease of the polarization.
The remaining features in Fig.~\ref{fig5} can be understood in similar terms.

\begin{figure}[ht]
\centerline{
\epsfig{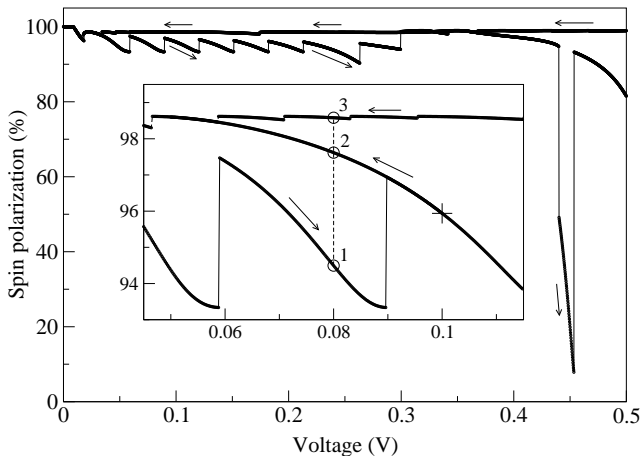}
}
\caption{Multistability of distinct polarization steady states within
the magnetic well of a $N=9$ MQW system.
The inset shows a blow-up of  three different steady states reached at $V=0.08$~meV.
State labeled 1 (3) is achieved by sweeping voltage up (down) from a high (low) initial bias.
State labeled 2 is obtained by sweeping voltage up to $V=0.1$
(marked with a cross) and then reversing the sweep direction.
}
\label{fig10}
\end{figure}

So far all results were calculated by sweeping voltages up.
Steady state solutions in the NDC bias voltage region are in general multistable.
We can obtain different solutions at a given bias voltage by evolving solutions following 
different histories,~\cite{kas94} for example by decreasing voltages from a high initial bias.
For a given voltage different values of the current with different density and spin
polarization profiles may be achieved.
To amplify effects, in studying this possibility,
we have set $\tau_{\rm sf}=10$~ns in a $N=9$ superlattice.
The change in $P$ with voltage is now so greatly increased
(see Fig.~\ref{fig10}) that even \emph{reversed} polarization can be observed 
at somewhat larger values of $\tau_{\rm sf}$
in the $E_1\rightarrow E_2$ resonant tunneling regime (not shown here).
This is a direct consequence of the emergence of dominant spin bottlenecks.\cite{jun98}
The alterations in polarization are more visible for voltages greater than the 
one corresponding to the $E_1\rightarrow E_2$ resonance, but 
we choose not to show them here since these involve transitions to higher excited energy levels,
close to the top of the barrier, where our model breaks down.
However, for sufficently high barriers it would be natural to obtain such behavior.

Fig.~\ref{fig10} shows three different values of spin polarization
in the magnetic well
which can be obtained at a particular bias voltage, depending on the sample
history, up sweep from zero voltage, down sweep from a high voltage, and 
up sweep to an intermediate voltage followed by down sweep.
We emphasize that this kind of hysteretic phenomena between magnetic states
is driven here by \emph{electric} fields.

\section{Conclusion}
\label{sec-con}
We have introduced and studied a simple model for growth direction non-linear 
transport in multiple quantum well systems containing magnetically doped layers.  Our 
analysis is based on a tunneling Hamiltonian expression for the current between
spin-polarized quantum wells and on a phenomenological expression for spin-relaxation
within quantum wells.  Numerical studies of this model 
show that it predicts rich behavior due to its non-linearity and due to 
the additional degrees of freedom introduced by spin-dependent transport.
Nonlinearity manifests itself in the formation of electric field domains when the
differential conductivity between neighboring layers is negative.
We find that the spin polarization of electrons in magnetic wells can change 
substantially when the system jumps between different branches of the $I$--$V$ curve. 
When current flows finite electron spin-polarization extends from the 
magnetic quantum wells to non-magnetic quantum wells.
When large equilibrium spin splitting, the $I$--$V$ curve is strongly affected by the appearance of
extra branches, due to tunneling into and out of spin polarized well subbands and these 
effects become more and more prominent when the characteristic time for spin-relaxation 
is longer.

The effects addressed in this paper could be investigated experimentally
by studying transport properties and by studying the polarization dependence 
of inter-band optical absorption and photoluminescence in MQW systems containing 
magnetically doped layers.    
The predicted sensitivity of transport properties to external magnetic fields,
suggests that these systems could potentially be useful for magnetic field
sensors, most likely in geometries with a relatively small number of quantum
wells.  The sensitivity to external fields will be strongest at low temperatures
where the the Mn ions are easily polarized to produce the maximum 
equilibrium spin-splitting field.  To illustrate the sort of effects that we expect to 
occur, we have considered only relatively simple geometries
with a single magnetic layer. Other effects will occur in larger MQW systems with 
particular geometries.
In general there is considerable lattitude 
for designing the Mn density distribution in the MQW system to realize desired magnetoresistance
effects that could be described with the type of model we have introduced here. 

Analogs of the magnetotransport effects we discuss will also occur in ferromagnetic 
multiple quantum well systems, similar to the delta-doped layered (Ga,Mn)As systems
studied by Kawakami {\it et al.}~\cite{kaw00} These systems are ferromagnetic and 
the carriers are holes rather than electrons, leading to strain-sensitive 
spin-orbit-coupling induced magnetic anisotropy\cite{die00,abo01} and coercitivities.
These properties suggest 
a rich interplay between the hysteretic magnetoresistance effects common
in thin film itinerant electron magnets\cite{bai88,tan01} and the hysteretic effects
discussed here, which have their roots in electric field domain structures.\cite{ohn00}
Had ferromagnetism been taken into account in our model calculations,
by solving for $\Delta$ self-consistenly in Eq.~(\ref{eq-delta}), not only the electric
field domain structure but also 
the magnetic state configuration would have been sensitive to the bias voltage history.
Exploring these possibilities appears to be a promising avenue for future experimental
and theoretical work. 

\acknowledgements

One of us (D.S.) thanks the hospitality of Indiana University
and The University of Texas at Austin where most of this work was completed.
This work was supported by the Spanish DGES Grant No. PB96-00875,
by the European Union TMR Contract FMRX-CT98-0180, and by the Indiana 21st century fund,
the Welch Foundation and DARPA/ONR Award No. N00014-00-1-0951.
The authors acknowledge valueable assistance from Tomas Jungwirth.

\end{multicols}

\end{document}